\documentclass[prb,preprint,onecolumn,amsmath,amssymb]{revtex4}
\usepackage{graphicx}
\usepackage{dcolumn}
\usepackage{bm}
\usepackage{epsfig}
\usepackage{subfigure}

\makeatletter
\newcommand{\printfnsymbol}[1]{%
  \textsuperscript{\@fnsymbol{#1}}%
}
\makeatother

\usepackage{multirow,tabu}

\newcommand{\bi}{\begin{itemize}}
\newcommand{\ei}{\end{itemize}}
\newcommand{\be}{\begin{eqnarray}}
\newcommand{\ee}{\end{eqnarray}}
\newcommand{\beq}{\begin{equation}}
\newcommand{\eeq}{\end{equation}}
\newcommand{\beqn}{\begin{equation*}}
\newcommand{\eeqn}{\end{equation*}}
\newcommand{\bbmatrix}{\left( \begin{array}}
\newcommand{\eematrix}{\end{array} \right)}

\begin{document}
\title{Band Alignment in Quantum Wells from Automatically Tuned {DFT+$U$}}
\author{Grigory Kolesov$^{1,2}$\footnote{email: gkolesov@seas.harvard.edu}\printfnsymbol{3}, Chungwei Lin$^1$\footnote{email: clin@merl.com}\footnote{These authors contributed equally.},  
Andrew Knyazev$^1$, Keisuke Kojima$^1$, Joseph Katz$^1$, \\
Koichi Akiyama$^3$, Eiji Nakai$^4$, Hiroyuki Kawahara$^4$
} 
\affiliation{
$^1$Mitsubishi Electric Research Laboratories, 201 Broadway, Cambridge, MA 02139, USA \\
$^2$Harvard University, Cambridge, MA 02138, USA \\
$^3$Mitsubishi Electric Corporation, 8-1-1, Tsukaguchi Honmachi, Amagasaki, Hyogo, Japan, 661-8661 \\
$^4$Mitsubishi Electric Corporation, 4-1, Mizuhara, Itami, Hyogo, Japan, 664-8641
}
\date{\today}

\begin{abstract}

Band alignment between two materials is of fundamental importance for
multitude of applications. However, density functional theory (DFT)
either underestimates the bandgap - as is the case with local density
approximation (LDA) or generalized gradient approximation (GGA) - or is highly
computationally demanding, as is the case with hybrid-functional methods. The
latter can become prohibitive in electronic-structure calculations of
supercells which describe quantum wells. We propose to apply the DFT$+U$ method, with
$U$ for each atomic shell being treated as set of tuning parameters, to automatically
fit the bulk bandgap and the lattice constant, and then use thus obtained $U$
parameters in large supercell calculations to determine the band alignment. We
apply this procedure to InP/In$_{0.5}$Ga$_{0.5}$As,
In$_{0.5}$Ga$_{0.5}$As/In$_{0.5}$Al$_{0.5}$As and InP/In$_{0.5}$Al$_{0.5}$As
quantum wells, and obtain good agreement with experimental results.  Although
this procedure requires some experimental input, it provides both meaningful
valence and conduction band offsets while, crucially, lattice relaxation is
taken into account. The computational cost of this procedure is comparable to
that of LDA.  We believe that this is a practical procedure that can be useful
for providing accurate estimate of band alignments between more complicated alloys.

\end{abstract}

\maketitle

\section{Introduction}

Band alignment between two materials is crucial for many industrial
applications, such as light-emitting diodes and diode lasers
\cite{diode_laser}, field-effect transistors \cite{modfet},
photovoltaics \cite{solar_perovskite}, photocatalysts \cite{Scanlon-2013},
photon waveguides \cite{Kuo-2005} and others. The most common
theoretical approach used to determine the band alignment is density functional theory
(DFT) \cite{PhysRev.140.A1133}, which is usually adequate for qualitatively 
comparing different materials, but is unsatisfactory quantitatively. One
serious difficulty of DFT is that it underestimates the bandgap when using
standard local density approximation (LDA)  or generalized gradient
approximation (GGA)  exchange-correlation functionals.  When using GGA or LDA
to determine the band alignment, only the valence band offset (VBO) can be
directly determined by the calculation with acceptable accuracy; the conduction
band offset (CBO) is inferred from the experimental bandgap of the bulk
materials
\cite{PhysRevB.41.12106,Hybertsen-1991,vandewalle_martin_1987,lany_zunger}.
Using this approach, the CBO cannot be determined when the interface strain
changes the bandgap of a material. The precision of a bulk bandgap can be greatly
improved by using better approximations, such as the many-body GW approach
\cite{PhysRev.139.A796, zhu_louie_1991} or hybrid-functional DFT
\cite{Perdew-1996, hse_solids}. These calculations, however, require
significantly more computational resources than those required for LDA or GGA, so
that supercell calculations to determine the band alignment can become too time
consuming, and supercell relaxation is often out of reach.  DFT$+U$ is
a method where the exchange-correlation functional is corrected by a set of $U$
values which are applied to selected atomic orbitals \cite{anisimov,dudarev}. DFT$+U$
allows adjustment of the bulk bandgap to the experimental value by using $U$ values as
tuning parameters. This approach was explored, for example, in Ref.
\cite{lany_zunger}  with application to the band-alignment problem. The results are 
not satisfactory in that DFT$+U$, while making bandgap-fitting possible for a
fixed lattice structure, does not reproduce the proper structure of the material
when the structure is allowed to relax. The same authors also proposed a different
empirical approach using the non-local external potential \cite{lany_zunger_nlep} which
provides the orbital-dependent energy shift to correct the bandgap. 
Another promising approach is the use of meta-GGA functionals such as the modified
Becke-Johnson functionals \cite{BJ,BR,TB09,TB09-GW}, which are computationally
inexpensive, but provide better estimation of the bandgap.  While {\em ab
initio} methods such as GW, hybrid-functional and meta-GGA DFT offer significant
improvements over LDA/GGA-functional DFT they are still quite problematic to use
in practical calculations: GW is accurate only in its self-consistent
realization  \cite{faleevGW,shilfgaardeGW} for some materials and less accurate
for others \cite{chantisGW, kang_hybertsen}; the computational cost of this method
is prohibitive in supercell and lattice relaxation calculations. Both hybrid-functional and
meta-GGA DFT require tuning of the parameters of the functional (e.g. screening
length and fraction of exchange)  to the material to achieve sufficient
accuracy \cite{wadehra_HSE,gmitra,wang2013}, while large supercells and
lattice relaxation are still difficult with the former.

In this paper, we re-examine DFT$+U$ as a practical ``black-box'' method for the
determination of the band alignment between two semiconductors. The $U$ values
of the bulk material are determined completely automatically by an optimization
procedure which adjusts them until the calculation reproduces i) the
experimental bandgap \emph{and} ii) the lattice parameters. The same $U$ values are
then used in the superlattice calculations. This procedure is semi-empirical, in
the sense that some experimental inputs are needed. However, it takes the
interface strain into account and results in accurate VBO and CBO, while
using minimalistic basis sets and computational resources comparable to
those required by LDA or GGA functionals.  We apply this procedure to
In$_{0.5}$Ga$_{0.5}$As/InP (denoted as InGaAs/InP below) and In$_{0.5}$Ga$_{0.5}$As/In$_{0.5}$Al$_{0.5}$As (InGaAs/InAlAs) superlattices, with
varying InGaAs widths. All alloys studied here are lattice-matched to InP. The change of the band alignment is quantitatively
consistent with the reported experiments, and bandgaps of the full superlattice
are consistent with photo-luminescent (PL) measurements. 
The same procedure is applied to InP/InAlAs to test transitivity~\cite{transitivity}. 
The rest of the paper
is organized as follows. In Section II we describe the procedure to determine
the band alignment, including a summary of bulk experimental values. In Section
III we show our results for InP/InGaAs, InGaAs/InAlAs, and InP/InAlAs lattice calculations.
The comparison to the photo-luminescent measurements is shown and discussed. A
brief conclusion is given in Section IV.

\section{Method}

\subsection{Computational details}
In the main part of this work we used the SIESTA package \cite{siesta02}. The
pseudopotential input files were downloaded from the SIESTA website.  In and Ga
pseudopotentials were generated with $4d$ (In) and $3d$ (Ga) electrons included
in the valence.  We used single-$\zeta$ + polarization shell (SZP) basis sets,
which were optimized in a bulk setup (such as GaAs or InP) using the Optimizer tool
from the SIESTA package \cite{garcia2009optimal} with basis pressure equal to 0.2
GPa and the Perdew-Burke-Ernzerhof (PBE) DFT functional \cite{PBE}.  The
optimized SZP basis sets have been shown to have similar quality to a
generic double-$\zeta$ + polarization (DZP) basis \cite{garcia2009optimal}.
The spin-orbit coupling was not included in the calculations. This is done in order to reduce the computational cost and in order to avoid convergence difficulties while probing different sets of $U$ parameters. Thus we attempt to capture the essential features of the quantum wells, that is band alignment in the relaxed structures, with $U$ parameters only.
All geometry relaxations were performed using the conjugate gradient method.

For the bulk calculation we used conventional unit cells and $7\times7\times7$
Monkhorst-Pack k-point sampling. All materials considered in this work have
zinc-blend structures. For alloy materials such as InGaAs we
also used the conventional unit cell. The same unit cells were replicated to
construct the interface  supercells. We did not use virtual crystal approximation (VCA)
or coherent potential approximation (CPA)
\cite{Hybertsen-1991,PhysRevB.41.12106} in this work.

%

DFT$+U$ \cite{anisimov,lichtenstein_strong, dudarev} is a method which is in
principle close to the hybrid-functional approach in that it attempts to
address the electron-electron interaction problem of local DFT
functionals \cite{cococcioni2005,ivady2014}. In the DFT$+U$ approach an 
 atomic orbital-dependent $U$ correction is added to
the DFT Hamiltonian\cite{anisimov}. In the Dudarev spherically averaged
approach \cite{dudarev}, which was employed here, this results in an effective orbital-dependent potential:
\beq
	V^{\text{LDA}+U}_{jk} = V^{\text{LDA}}_{jk} + U \left[ \frac{1}{2}\delta_{jk} - \rho_{jk} \right],
\label{Vu}
\eeq
where $j,k$ are orbital indices  and $\rho$ is the electronic single-particle density matrix.
The parameter $U$ for each orbital can in principle be computed \emph{ab
initio} \cite{cococcioni2005,aryasetiawan2006,ivady2014,agapito2015} but in practice
is often fitted to reproduce experimental results such as the bandgap. Eq.
(\ref{Vu}) shows that for positive $U$ the energy levels are shifted up for
unoccupied orbitals and down for occupied ones.

\subsection{$U$ optimization\label{simplex}}
In this work we used the DFT$+U$ approach where $U$ values were fitted in a systematic way. Given
a bulk crystal structure we enable $U$ for each valence atomic orbital, except
for semicore orbitals such as $4d$ in In, which are completely filled and lie
very deeply in the valence band of the materials we study here.  We then apply
the simplex method as implemented in the Optimizer tool from the SIESTA package to
minimize an objective function:

\beq
    f\left(\mathbf{U}\right) = w_g \left[ E_g\left(\mathbf{U}\right) - E^{\text{exp}}_g \right]^2 +
                        w_a \sum_{i=1}^3 \left[\mathbf{a}_i(\mathbf{U}) - \mathbf{a}_i^{\text{exp}}  \right]^2.
\label{fopt}
\eeq
Here $\mathbf{U}$ denotes set of all values of $U_j$, $j$ being a combined index
for an atomic species and $n$,$l$ quantum numbers, $E_g$ and $\mathbf{a}_i$ denote
bandgap and lattice vectors respectively and superscript ``exp'' denotes
experimental values. $w_g$ and $w_a$ are weights, which we chose to be $0.33$
eV$^{-2}$ and $0.67$ \AA$^{-2}$. For a given $\mathbf{U}$, the full lattice relaxation
followed by a bandgap computation is performed. 

We would like to make a few remarks. i) The minimization is deemed sufficient when $f(\mathbf U )\lesssim 10^{-3}$
because of experimental uncertainties. ii) Here $\mathbf{U}$ are
treated as free parameters, which are not only aimed at correcting deficiencies
of the PBE functional but also serve as a finite basis set
correction \cite{kulik_basisU}. 
Thus $U$'s could in
principle be  negative, although in this work we restrict them to be positive.
iii) The optimization is performed on the bulk unit cell and is
computationally inexpensive, typically taking several hours on four CPU cores.

\subsection{Determination of the band alignment}

We consider band alignments between InGaAs and InP, and between
InGaAs and InAlAs. In both cases
InGaAs is the ``well'' material which has a bandgap of 0.82eV;
InP and InAlAs serve as the ``barrier'' materials whose
bandgaps are around 1.4 eV. All three materials have lattice constants of
5.86\AA, lattice matched to InP. The band diagram of a quantum well or a
superlattice is illustrated in Fig.~\ref{fig:InGaAs_InP}(a).

To determine the band alignment, superlattices of (InGaAs)$_n$/(InP)$_{10}$,
(InGaAs)$_n$/(InAlAs)$_{10}$, and (InAlAs)$_{10}$/(InP)$_{10}$ are used, with the conventional zinc blende unit
cell serving as the basic building block. As illustrated in
Fig.~\ref{fig:InGaAs_InP}(a), the supercell has a period of $1\times 1 \times
(10+n)$, with the stacking direction defined as $z$. 
Because all three materials have almost the same lattice constant which is reproduced
in our bulk calculations with optimized $U$ parameters, we fix the in-plane
lattice constant to that of bulk InP  and allow only  relaxation of the
supercell in the $z$-direction and relaxation of the ionic positions in our calculations. 
As the projected density of states (DOS) recovers its bulk profile
away from the interface, the band alignment is determined by the projected DOS
in the middle of InGaAs, InP, and InAlAs
respectively.

%

\subsection{Summary of bulk experiments\label{empformulas}}

We conclude this section by summarizing the experimental results of two classes of III-V zinc-blend  alloys.
The first class is Ga$_x$In$_{1-x}$As$_y$P$_{1-y}$, whose lattice constant is given by \cite{doi:10.1063/1.329882, doi:10.1063/1.90455}
\beq
a_{\text{GaInAsP}}(x,y) = 5.8696-0.4184 x + 0.1894 y + 0.0130 xy.
\label{eqn:lattice}
\eeq 
The bulk bandgaps are \cite{doi:10.1063/1.90455}
\beq
E_{\text{GaInAsP}}(x,y) = 1.35 + 0.668 x -1.17 y + 0.758 x^2 + 0.18 y^2 -0.069 xy -0.322 x^2 y + 0.03 xy^2
\label{eqn:bandgap}
\eeq 
Eq.~\eqref{eqn:lattice} and ~\eqref{eqn:bandgap} are room-temperature results. When lowering the temperature, the bandgap becomes larger and the lattice constant smaller. For example, InP at 4K has a bandgap  around 1.45 eV and a lattice constant around 5.85\AA. 

The second class of alloys is In$_{1-x-y}$Ga$_x$Al$_y$As. The physical quantities can be parametrized as \cite{Minch-1999}
\beq 
\begin{split}
P(\mbox{In$_{1-x-y}$Ga$_x$Al$_y$As}) &= P(\mbox{InAs})(1 -x -y)+ P(\mbox{GaAs}) x + P(\mbox{AlAs})y 
\end{split}
\eeq 
Using the bulk data summarized in Ref.~\cite{Zhang:11}, the lattice constants of GaAs,  GaP,  InAs, and AlAs are respectively 5.6533\AA,  5.4505\AA, 6.0584\AA, and 5.660\AA. The lattice constant of this class of alloys is therefore parameterized as
\beq 
\begin{split}
a_{\text{InGaAlAs}}(x,y) &= 6.0584 (1 -x -y) + 5.6533 x + 5.660 y\\
&= 6.0584 -0.4051 x - 0.3984 y. 
\end{split}
\eeq 
The bulk bandgap is obtained from 
\beq 
\begin{split}
E_{\text{InGaAlAs}} (\mbox{In$_{1-x-y}$Ga$_x$Al$_y$As}) &= 0.36 + 2.093y + 0.629x + 0.577 y^2 \\
&+ 0.436x^2 + 1.013xy - 2.0 xy( 1 - x - y )eV.
\end{split}
\label{eqn:bandgap_al_general}
\eeq
For alloys that are lattice matched to InP (5.86 \AA) where $x+y = 0.47$, i.e., In$_{0.53}$Ga$_{0.47-y}$Al$_y$As,  the bandgap fitted from Ref.~\cite{doi:10.1063/1.93537}
\beq
E(y) = 0.76 \pm 0.04 + (1.04 \pm 0.10) y + (0.87 \pm 0.13) y^2.
\label{eqn:bandgap_al}
\eeq 
Directly using Eq.~\eqref{eqn:bandgap_al_general}, we get $E(y) = 0.7519 + 1.0321 y + 1.06 y^2$. It appears that the coefficient of $y^2$ is not consistent between these two expressions. However since $y<0.47$, the error is at most $(1.06-0.87)\times 0.47^2 = 0.042$ eV, which sets the uncertainty in our calculations. The experimental results summarized here are used to optimize the $U$ values in the DFT$+U$ functional.

\section{Results}


\subsection{Optimized U values}

\begin{table}[ht]
\begin{tabular}{c | c | c | c | c | c | c  }  
    Material & $E_g-E_g^{\text{exp}}$ (eV) & $a - a^{\text{exp}}$ (\AA) & Species  &   $n$ quantum number &  $U_s$ (eV) & $U_p$ (eV)  \\ \hline
    \multirow{ 3}{*}{In$_{0.5}$Ga$_{0.5}$As} & \multirow{ 3}{*}{-0.01}  & \multirow{ 3}{*}{-0.02}  & In & 5 & 0.04 & 7.43 \\
\cline{4-7}
                                            & & &  Ga & 4 & 0.00 & 4.23 \\
\cline{4-7}
                                            & & &  As & 4 & 0.00 & 0.00 \\ 
\hline
\hline
    \multirow{ 2}{*}{InP}                    & \multirow{ 2}{*}{-0.03}  & \multirow{ 2}{*}{-0.02}  &  In & 5 & 0.00 & 4.23 \\
\cline{4-7}
                                             & & & P  & 3 & 0.00 & 0.48 \\
\hline
\hline
    \multirow{ 3}{*}{In$_{0.5}$Al$_{0.5}$As} & \multirow{ 3}{*}{0.00}  & \multirow{ 3}{*}{0.00} & In & 5 & 0.01 & 3.31 \\
\cline{4-7}
                                             & & &  Al & 3 & 0.41 & 2.80 \\
\cline{4-7}
                                             & & &  As & 4 & 0.03 & 0.15 \\
\end{tabular}
\caption{ $U$ parameters optimized for best fit to experimental values of bandgap ($E_g$) and lattice constant ($a$) (Section \ref{simplex}).  $U_s$ denotes the value of $U$ applied to the $s$-shell of the corresponding atom, while $U_p$ denotes $U$ values applied to $p$-shells. With the values of $U$ presented here all angles are 90$^\circ$ in the optimized conventional unit cells. 
}
\label{table:UValues}
\end{table}

The $U$ values for InP, InGaAs, and
InAlAs are given in Table~\ref{table:UValues}. As
described in Section \ref{simplex}, these $U$ values are computationally
optimized to fit both experimental bulk lattice structure and bandgap.
It can be seen that $U$ values of the same species strongly depend on the material.
For example for the $5p$-shell of the In atom, the value of $U_p$  is 7.43,
4.23 and 3.31 eV in InGaAs, InP and InAlAs respectively.  This is expected,
because the value of $U$ incorporates screening effects \cite{anisimov} and
thus depends on the environment of the atom.

There are two other trends clearly visible in Table \ref{table:UValues}.
First, the values $U_s$ for all species are small or nearly zero. Second, the
$U$ values for anionic atoms are much smaller then $U$ for cationic atoms. This
indicates that both the optimization of the geometry and the bandgap is largely
controlled by $p$-states of the cationic atoms, which constitute the largest part
of the conduction band and a smaller but not insignificant part of the valence
band.


\subsection{InGaAs/InP, InGaAs/InAlAs, and InAlAs/InP }

\begin{figure}[htp]
\includegraphics[width = 1.0\textwidth]{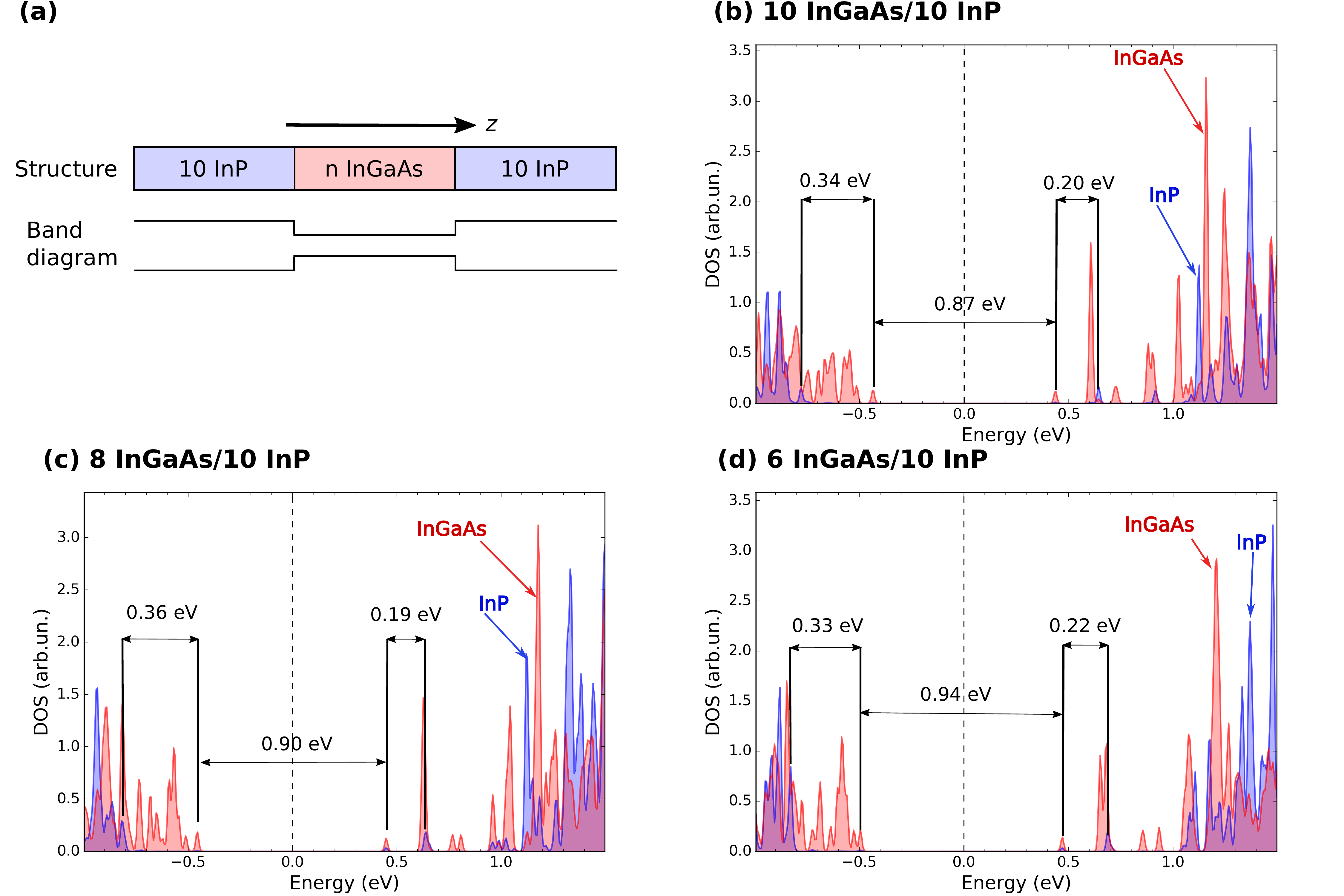}
\caption{(a) Illustration of the structure and the band diagram of the
superlattice used to determine the band alignment between InP and InGaAs. (b)
Projected DOS for (InGaAs)$_{10}$/(InP)$_{10}$. (c) Projected
DOS for (InGaAs)$_8$/(InP)$_{10}$.  (d) Projected DOS for
(InGaAs)$_6$/(InP)$_{10}$.  The intervals show, from left to
right, VBO, bandgap and CBO.}
\label{fig:InGaAs_InP}
\end{figure}

Fig.~\ref{fig:InGaAs_InP}(b)-(d) show the computed projected DOS for
(InGaAs)$_n$/(InP)$_{10}$ with $n=10,8,6$. The bandgap
increases as the quantum well width, characterized by $n$, decreases due to the
increasing quantum confinement. The computed bandgap is in excellent agreement with
the PL experiments, as summarized in Table~\ref{table:Comparison_1}. The band
alignment is about the same for all three quantum well widths: the VBO is
around 0.35 eV whereas the CBO is around 0.20 eV. This falls within the range
of experimental values, where the VBO and CBO were measured at $\sim$0.35 eV and $\sim$0.22 
eV on average (Refs.~\cite{0268-1242-1-1-003, vurgaftman2001, adachi_book} and references therein). 


\begin{figure}[htp]
\includegraphics[width = 1.0\textwidth]{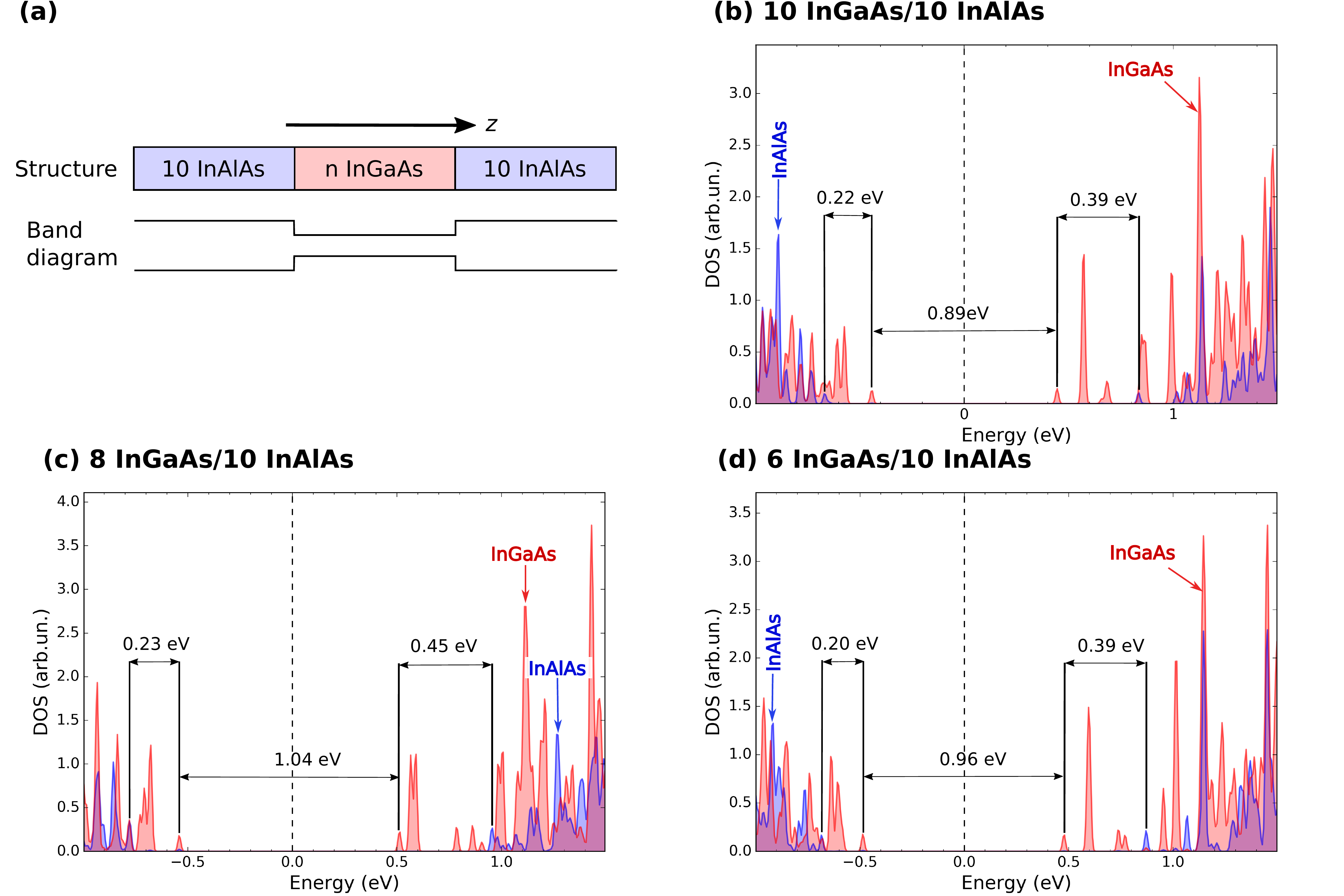}
\caption{(a) Illustration of the superlattice structure used to determine the band alignment between InAlAs and InGaAs.  (b) Projected DOS for (InGaAs)$_{10}$/(InAlAs)$_{10}$. (c) Projected DOS for (InGaAs)$_8$/(InAlAs)$_{10}$.  (d) Projected DOS for (InGaAs)$_6$/(InAlAs)$_{10}$. The intervals show, from left to
right, VBO, bandgap and CBO. }

\label{fig:InGaAs_InAlAs}
\end{figure}

Fig.~\ref{fig:InGaAs_InAlAs}(b)-(d) show the computed projected DOS for
(InGaAs)$_n$/(InAlAs)$_{10}$ with $n=10,8,6$.
The bandgap also opens up with decreasing $n$ due to the stronger quantum
confinement.  The band alignment is roughly the same for all three quantum well
widths: the VBO is around 0.22 eV whereas the CBO is around 0.41 eV. In
Refs.~\cite{doi:10.1063/1.337262,vurgaftman2001} In$_{0.53}$Ga$_{0.47}$As (0.78
eV)/In$_{0.52}$Al$_{0.48}$As (1.44 eV, 5.85\AA) is shown to have a VBO and CBO 
on average of 0.22 eV and 0.50 eV respectively; the VBO is about the same,
whereas the CBO is $\sim$0.1 eV larger than the computed values. We note that one of
the lower experimental values reported is $0.47\pm0.03$ eV \cite{lopez1993} and
our lower computed value is consistent with the lower fraction of In 
used in the computation (0.5 vs lattice-matching 0.53 in experiment) \cite{lopez1993}.

As InP and InAlAs have similar bandgaps and lattice constants,
our calculations show that InGaAs/InP has the larger VBO and
the smaller CBO, whereas InGaAs/InAlAs has the larger
CBO and the smaller VBO. This trend is consistent with the experiments.  Generally,
the VBO/CBO values depend only weakly on the quantum well width. The bandgap,
however, displays an observable dependence on quantum well width, and will be
discussed in Section \ref{PLexp} in terms of photoluminescent measurements. 

\begin{figure}[htp]
\includegraphics[width = 0.8\textwidth]{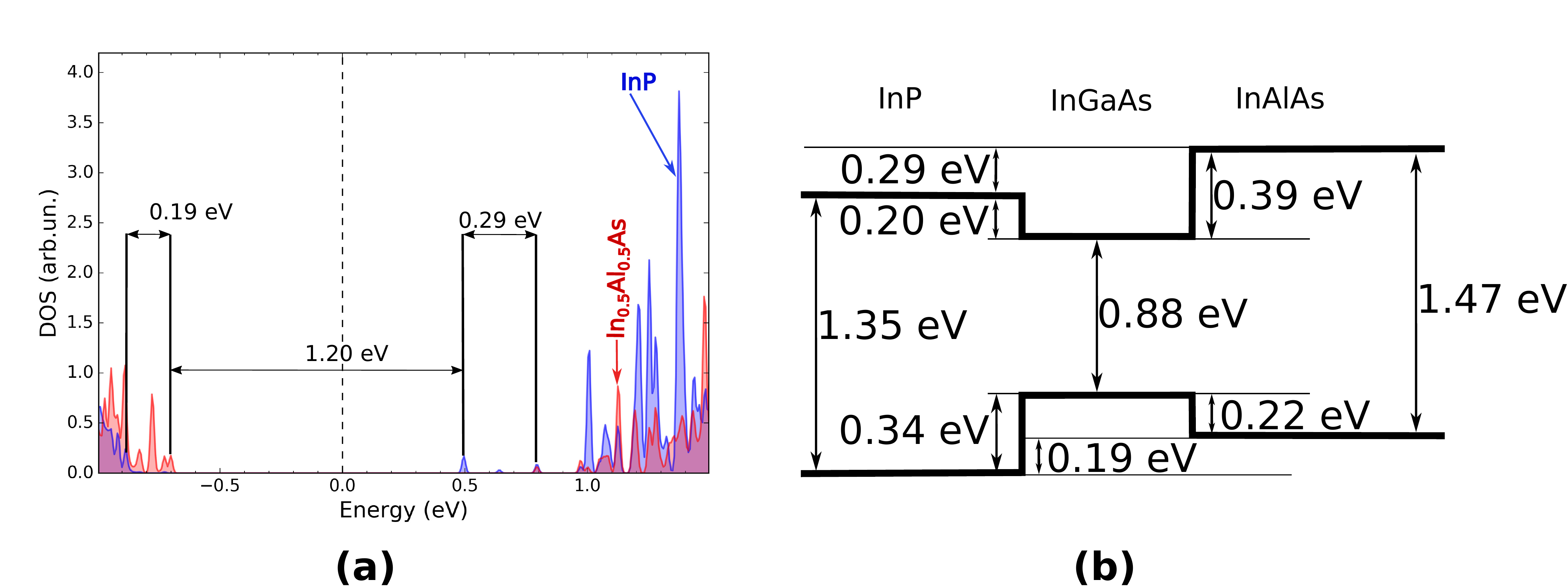}
\caption{
(a) Projected DOS for InAlAs$_{10}$/(InP)$_{10}$. The intervals show, from left to
right, VBO, bandgap and CBO. 
(b) The combined band diagram InP/InGaAs/InAlAs. Values of VBO and CBO are indicated. 
The VBO's and CBO's of different interfaces satisfy the transitivity rule within $\sim 0.1$ eV.
}
\label{fig:InAlAs_InP}
\end{figure}


To check the transitivity of the proposed procedure, we compute the band
alignment using (InAlAs)$_{10}$/(InP)$_{10}$, as shown in
Fig.~\ref{fig:InAlAs_InP}(a). Treating InP as the quantum well, the CBO and VBO
are respectively around 0.29 eV and -0.19 eV. These values are at about the average
of the experimental values \cite{adachi_book, vurgaftman2001}.
Fig.~\ref{fig:InAlAs_InP}(b) shows the combined band diagram of all three
interfaces InP/InGaAs/InAlAs. The VBOs and
CBOs of these three materials are transitive within $\sim 0.1$ eV. This degree
of non-transitivity agrees with experiment~\cite{vurgaftman2001}.

\subsection{Comparison to photoluminescent measurements\label{PLexp}}

\begin{figure}[htp]
\includegraphics[width = 0.8\textwidth]{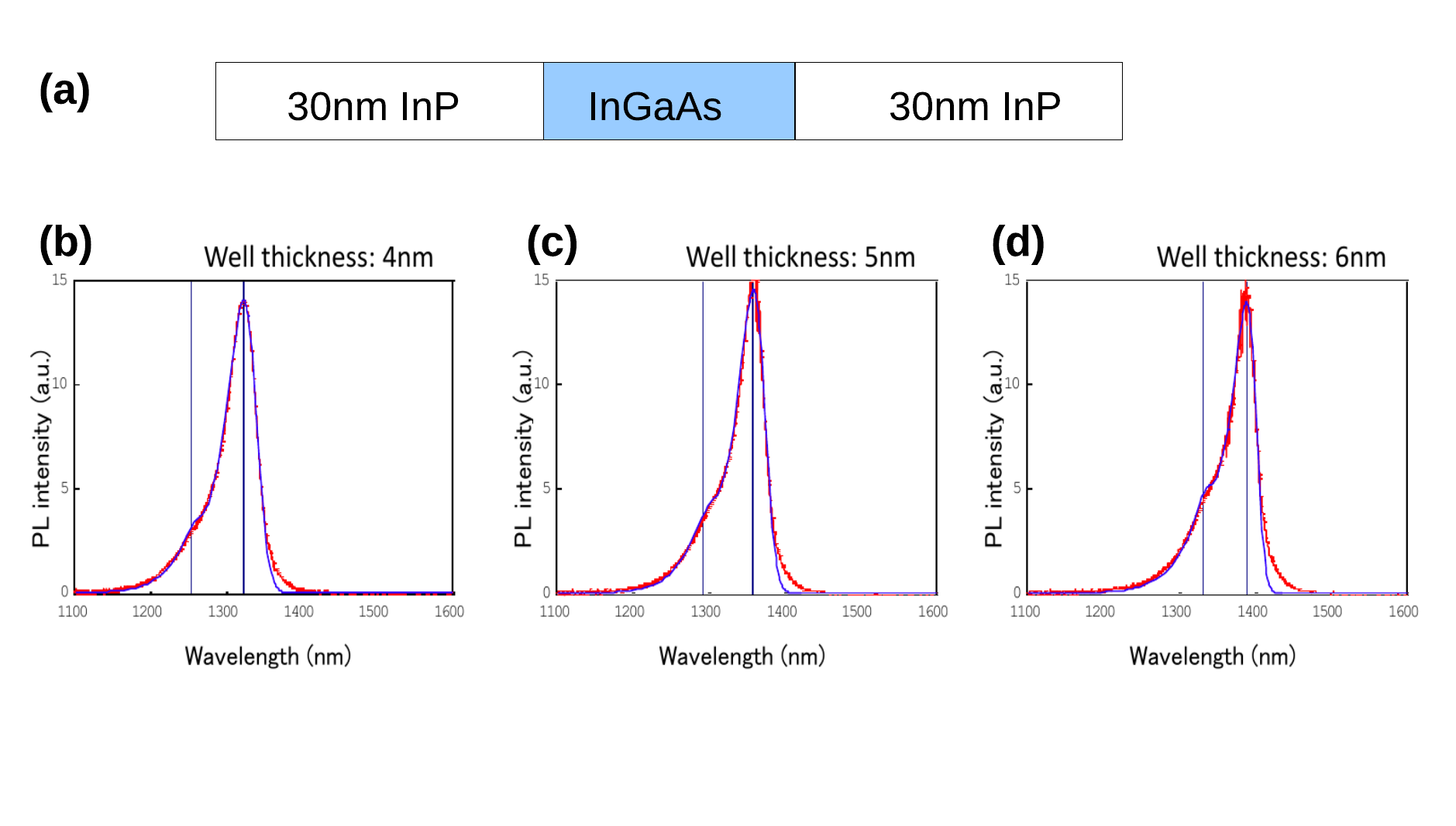}
\caption{(a) Illustration of the superlattice structure used for PL measurement. The barrier material InP is 30nm thick. (b)-(d) 
The PL measurement for 4nm In$_{0.53}$Ga$_{0.47}$As (b), 5nm In$_{0.53}$Ga$_{0.47}$As (c), and 6nm In$_{0.53}$Ga$_{0.47}$As (d).
The red curves are measurements, and blue curves are Gaussian fit. The first (lowest) peak values are summarized in Table~\ref{table:Comparison_1}.  }
\label{fig:InGaAs_InP_Expt}
\end{figure}

\begin{table}[ht]
\begin{tabular}{l  c   c   c  }  
  Quantum well  & Bandgap (theory) & PL measurement & Experimental well width \\ \hline
6InGaAs/10 InP  & 0.94 eV  & 0.94 eV & 4nm   \\ 
8 InGaAs/10 InP  & 0.90 eV  & 0.91 eV & 5nm   \\ 
10 InGaAs/10 InP & 0.87 eV  & 0.89 eV & 6nm   
\end{tabular}
\caption{ The bandgap of InGaAs/InP, as a function of quantum well width.
}
\label{table:Comparison_1}
\end{table}


To further test the calculations, we prepared quantum wells of 4nm, 5nm and 6nm InGaAs as well as 30nm InP, and perform the PL measurements. 
 The superlattice is grown using the standard MOCVD (Metal-Organic Chemical Vapour Deposition) method. The PL experiments were carried out at 300 K. 
The results are shown in Fig.~\ref{fig:InGaAs_InP_Expt}. The Gaussian fits imply that the PL spectra display at least two peaks, which we interpret as the heavy hole and light hole splitting. The observed lowest-energy peak corresponds to the bandgap of the quantum well, and is summarized in Table~\ref{table:Comparison_1}. As the lattice constant is 5.86~\AA,  InGaAs wells of width 4nm, 5nm, 6nm are close to 6, 8, 10 InGaAs unit cells. The computed bandgaps are also given in Table~\ref{table:Comparison_1}, and good agreement is seen.

%
\section{Conclusion}

In this paper, we demonstrate that DFT calculations using DFT$+U$ can be an
efficient way to determine the band alignments between two alloys. The full
procedure can be divided into two steps. The first step is to determine $U$
values of a bulk alloy by automatically optimizing atomic orbital-specific
values of $U$ so that the experimental bandgap and the lattice constant agree
with the values obtained in the simulation.  The second step is to use 
these fitted $U$ values in a superlattice calculation (with lattice relaxation),
and the valence and conduction band offsets are then determined from the
projected DOS away from the interface. We apply this procedure to
InGaAs/InP, InGaAs/InAlAs, and InAlAs/InP,
and are able to obtain both VBOs and CBOs consistent with experiments. 
The degree of non-transitivity $\sim 0.1$ eV in the calculated band alignments is in agreement with experiment.
In addition the computed quantum-well width-dependent bandgaps of InGaAs/InP are in excellent
agreement with the photoluminescent measurements.  The proposed method is
semi-empirical, because optimization of $U$ values requires knowledge of
experimental bandgaps and lattice constants. However, it provides meaningful
valence and conduction band offsets between two alloys, with the interface
strain taken into account. For many semiconductor alloys the experimental data
are available for at least 3 compositions, that is for $x=1,0$ and $0.5$ in
the A$_x$B$_{1-x}$C alloy. Because empirical composition-bandgap dependencies
(section \ref{empformulas}) are quadratic, it seems plausible that the set of $U$
values can be likewise interpolated by a quadratic polynomial.  The use of
a compact numerical atomic orbital basis sets as implemented in SIESTA package
makes this method quite lightweight, amenable to large (200+ atoms) supercell
computation on a single workstation.  Because lattice relaxation is taken into
account, the proposed procedure can serve as a practical method to explore the
band alignments between complicated alloys.  

\subsection*{Acknowledgement}
We thank Dr. Gilles Z\'erah and Prof. Efthimios Kaxiras for several insightful discussions. The photoluminescent measurements were performed in Amagasaki, Japan.




\begin{thebibliography}{48}
\bibitem[{tra()}]{transitivity}
\bibinfo{note}{The transitivity rule refers to the following: for three
  semiconductors, A, B, and C, the band offset at the heterojunction A/B can be
  deduced from the band offsets at the heterojunctions A/C and C/B.}
\end{thebibliography}
\end{document}